\documentclass{aastex}
\usepackage{spr-astr-addons}
\usepackage{url}
\usepackage{amsmath}
\usepackage{amsfonts}
\usepackage{amssymb}

\usepackage{graphicx}
\usepackage{subfigure}

\RequirePackage{color}

\begin{document}

\title{ Strong and weak gravitational field in $R+\mu^4/R$ gravity}
\slugcomment{Not to appear in Nonlearned J., 45.}
%% Running heads
\shorttitle{Strong gravitational field in ...}

\shortauthors{Kh. Saaidi et al.}
\author{$^a$Kh. Saaidi\altaffilmark{1}}\and\author{$^{a}$A. Vajdi}  \and \author{ $^a$S. W. Rabiei} \and \author{ $^b$A. Aghamohammadi}
\and  \author{ $^a$H. Sheikhahmadi}
 \affil{$^a$Department of Physics, Faculty of Science, University of
 Kurdistan,  Sanandaj, Iran.}
 \altaffiltext{1}{ksaaidi@uok.ac.ir.}
 \affil{$^b$Department of Physics,  Sanandaj Branch, Islamic Azad
 University, Sanandaj, Iran}
%\altaffiltext{2}{ksaaidi@uok.ac.ir.}
%\altaffiltext{3}{Third Alternate Affilation.}

\begin{abstract}
We introduce a new approach for investigating
the weak field limit of vacuum field equations in $f(R)$ gravity and we find the weak field limit of  $f(R)=R+\mu ^4/R$ gravity. Furthermore,
we study the strong gravity regime in $R+\mu^{4}/R$ model of $f(R)$ gravity. We show the existence of strong gravitational field in vacuum for such model.
 We find out in the limit $\mu\rightarrow 0$ , the weak field limit and
 the strong gravitational field can be regarded as a perturbed Schwarzschild metric.
\end{abstract}


%\section*{}
%\label{sec:intro}
\section{Introductions}
Observations on supernova type Ia \citep{f1a, f1b}, cosmic microwave
background \citep{f2}
 and large scale structure \citep{f3}, all
indicate that the expansion of the universe is not proceeding as
predicted by general relativity, if the universe is homogeneous,
spatially flat, and filled with relativistic matter. An interesting
approach to explain the positive acceleration of the universe is
$f(R)$ theories of gravity which generalize the geometrical part of
Hilbert-Einstein lagrangian \citep{c3, c4, c6, p, f10, r14, r8,
kla}. One of the initiative $f(R)$ models supposed to explain the
positive acceleration of expanding universe has $f(R)$ action as
$f(R)=R-\mu^4/R$ \citep{c4}. After proposing the $f(R)=R-\mu^4/R$
model, it was appeared this model suffer several problems. In the
metric formalism, initially Dolgov and Kawasaki discovered the
violent instability in the matter sector \citep{r3}. The analysis of
this instability generalized to arbitrary $f(R)$ models
\citep{r4,r5} and it was shown than an $f(R)$ model is stable if
$d^{2}f/dR^{2}>0$ and unstable if $d^{2}f/dR^{2}<0$. Thus we can
deduce $R-\mu^4/R$ suffer the Dolgov-Kawasaki instability but this
instability removes in the $R+\mu^4/R$ model, where $\mu^{4}>0$.
Furthermore, one can see in the $R-\mu^4/R$ model the cosmology is
inconsistent with observation when non-relativistic matter is
present. In fact there is no matter dominant era \citep{r6a, r6b,
r8}. However, the recent study shows the standard epoch of matter
domination can be obtained in the $R+\mu^4/R$ model \citep{r8}.\par
It is obvious that a viable theory of gravity must have the correct
newtonian limit. Indeed a viable theory of $f(R)$ gravity must pass
solar system tests. After the $R-\mu^4/R$ was suggested as the
solution of cosmic-acceleration puzzle, it has been argued that this
theory is inconsistent with solar system tests \citep{a5}. This
claim was based on the fact that metric $f(R)$ gravity is equivalent
to  $\omega=0$ Brans-Dicke theory, while the observational
constraint is $\omega>40000$. But this is not quite the case and it
is possible to investigate the spherical symmetric solutions of
$f(R)$ gravity without invoking the equivalence of $f(R)$ gravity
and scalar tensor theory \citep{p, New2, r14, v, r12, r13, rr13, rr14}. It
has been shown that
 some $f(R)$ models accept the Schwarzschild-de Sitter space-time as a spherical
 symmetric solutions of field equation\citep{v}. Hence $R-\mu^4/R$
 model has a Schwarzschild-de Sitter solution with constant curvature
 as $R=\sqrt{3 \mu^{4}}$ where this is not the case in  $R+\mu^4/R$
 model.\par
In this paper we study the $R+\mu^{4}/R$ model of $f(R)$
gravity. We find the static spherically symmetric solution of
vacuum field equation in both weak field limit and strong gravity regime,
moreover, the weak field analysis can be expanded on $f(R)$ models
of the form $f(R)=R+\epsilon h(R)$.
%=======================================================================================================================
\section{Weak field limit}
In this section we investigate the weak field solution of vacuum field
equation in $f(R)$ theories of gravity. We are interested in model
of the form $f(R)=R+\epsilon h(R)$, with $\epsilon$ an adjustable
small parameter. The motivation for discussing these models is that
the nonlinear curvature terms that grow at low curvature can lead to
the late time positive acceleration, but during the standard matter
dominated epoch, where the curvature is assumed to be relatively
high, could have a negligible effect.\par The field equations
for these models are
\begin{eqnarray}\label{eq1}
G_{\mu\nu}=-\epsilon\left[G_{\mu\nu}+g_{\mu\nu}\Box
-\nabla_{\mu}\nabla_{\nu}+ {g_{\mu\nu}\over 2}\right.
\nonumber
\\
\left.\times\left(R- {h(R) \over
\varphi(R)}\right)\right]\varphi(R)+kT_{\mu \nu},
\end{eqnarray}
where $\varphi(R)=dh(R)/dR$. Contracting the field equation we
obtain
\begin{eqnarray}\label{eq2}
R=\epsilon\left[R-\frac{2h(R)}{\varphi(R)}+3\Box\right]\varphi(R)-kT.
\end{eqnarray}
Where for the vacuum $T_{\mu \nu}, T=0$. If $\epsilon = 0$ the above equations reduce to Einstein equation.
Hence we suppose $G_{\mu \nu}$ and $R$ in the r.h.s of
Eqs.(\ref{eq1},\ref{eq2}) can be neglected for small values of
$\epsilon$. Furthermore if the condition $\mathop {\lim }\limits_{R
\to 0} \left[ {h(R)/\varphi (R)} \right] = 0$ is satisfied we can
neglect this term too. Neglecting these terms leads to the following
equations
\begin{eqnarray}\label{eq3}
G_{\mu\nu}=-\epsilon\left[g_{\mu\nu}\Box
-\nabla_{\mu}\nabla_{\nu}\right]\varphi(R),
\end{eqnarray}
and
\begin{eqnarray}\label{eq4}
R=\epsilon 3\Box \varphi(R).
\end{eqnarray}
The analysis of spherically symmetric solution can be carried out
using schwarzschild coordinate
\begin{eqnarray}\label{eq5}
 ds^{2}=-A(r)dt^{2}+B(r)^{-1}dr^{2}+r^{2}d\Omega^{2}.
\end{eqnarray}
In the weak field limit approximation the metric deviates slightly
from the Minkowski metric, so we can write
\begin{eqnarray}\label{eq6}
&& A(r)=1+a(r),
\nonumber
\\
&&B(r)=1+b(r),\nonumber
\\
&&\mid a\mid,\mid b\mid \ll 1.
 \end{eqnarray}
When solving the field equations(\ref{eq3},\ref{eq4}) we will keep
only terms linear in the perturbations $a(r)$, $b(r)$. Hence
equations (\ref{eq3},\ref{eq4}) leads to
\begin{eqnarray}\label{eq7}
\frac{a^{\prime}}{r}+\frac{b}{r^{2}}=-\epsilon\frac{2}{r}\frac{d\varphi(R)}{dr}\nonumber\\
\frac{b^{\prime}}{r}+\frac{b}{r^{2}}=-\epsilon\nabla^{2}\varphi(R),
 \end{eqnarray}
 and
 \begin{eqnarray}\label{eq8}
R=3\epsilon\nabla^{2}\varphi(R).
\end{eqnarray}
where $(\prime)$ indicates a derivation with respect to $r$.

\subsection{$f(R)=R^{1+\epsilon}$} %====================================================== f(R)
This model is considered in \citep{p}. It is shown that this model
has an exact spherically symmetric vacuum solution and regarding the
general line-element in Eq.(\ref{eq5}), it may be written as
\begin{eqnarray}
 A(r)&=&r^{2\epsilon(1+2\epsilon)/(1-\epsilon)}+c ~r^{-(1-4\epsilon)/(1-\epsilon)}\nonumber,\\
B(r)&=&\frac{(1-\epsilon)^{2}}{(1-2\epsilon+4\epsilon^{2})(1-2\epsilon-2\epsilon^{2})}
\nonumber
\\
&&\times\left(1+   c~r^{-(1-2\epsilon+4\epsilon^{2})/(1-\epsilon)}
\right)\nonumber,
 \end{eqnarray}
where $c$ is a constant. In the limit $\epsilon\rightarrow 0$, these
solutions become
\begin{eqnarray}\label{eq9}
 ds^{2}&=&-\left(1+2\epsilon\ln r+\frac{c}{r}\right)dt^{2}+
 \left(1+2\epsilon+\frac{c}{r}\right)^{-1}dr^{2}
 \nonumber
 \\
 &&+r^{2}d\Omega^{2}.
 \end{eqnarray}
because we seek the weak field limit, in above equation we assume $c/r\ll1$. \par
Since we are interested in the limit $\epsilon\rightarrow0$, we may
expand $f(R)=R^{1+\epsilon}$ around $\epsilon=0$. Then we have
\begin{eqnarray}
 f(R)&=&R+\epsilon R \ln  R, \nonumber\\
 h(R)&=&R \ln  R,\nonumber\\
 \varphi(R)&=&1+\ln R\nonumber .
 \end{eqnarray}
It is clear that $h(R)$ satisfies the condition
\begin{equation}\nonumber
\mathop {\lim }\limits_{R \to 0} \left[ {h(R)/\varphi (R)} \right]=
0.
\end{equation}
 Inserting $\varphi(R)$ in the trace equation (\ref{eq8}), the  Ricci scalar is
obtained as
\begin{eqnarray}\label{eq10}
 R=-\frac{6\epsilon}{r^{2}}.
 \end{eqnarray}
 Then we arrive at the solutions of Eq.(\ref{eq7})
\begin{eqnarray}\label{eq11}
 a=\frac{c}{r}+2\epsilon\ln r,& &b=\frac{c}{r}+2\epsilon,
 \end{eqnarray}
where $c$ is a constant. We can see our solutions are in agreement
with the exact solutions (\ref{eq9}). Also one can check neglecting
$R$, $G_{\mu\nu}$ and $h(R)/\varphi(R)$ in Eq.(\ref{eq1}, \ref{eq2})
is reasonable.

%----------------------------------------------------------------------------------
\subsection{$f(R)=R+\epsilon \ln R$}
For this model $\varphi(R)=1/R$. Solving trace equation (\ref{eq8})
and field equations (\ref{eq7}) we obtain
\begin{eqnarray}\label{eq15}
R=\frac{\sqrt{6\epsilon}}{r},
\end{eqnarray}
and
\begin{eqnarray}\label{eq16}
a=b=-\frac{2M}{r}-\sqrt{\frac{\epsilon}{6}}r.
\end{eqnarray}
where $M$ is a constant. Therefore the space time metric for empty
space in this model is
\begin{eqnarray}\label{eq17}
 ds^2&=&-\left(1-\frac{2M}{r}-\sqrt{\frac{\epsilon}{6}}r\right)dt^2
 \nonumber
 \\
 &&+\left(1-\frac{2M}{r}-\sqrt{\frac{\epsilon}{6}}r\right)^{-1}dr^2 + r^2d\Omega^2.
 \end{eqnarray}
 We can see, the generalized Newtonian
 potential  is
\begin{eqnarray}\label{eq18}
 \Phi_G=-\frac{M}{r}-{1\over 2}\sqrt{\frac{\epsilon}{6}}r.
 \end{eqnarray}
This generalized gravitational potential has two terms. The first
term is the standard Newtonian potential and the second term make a
constant acceleration, $+\sqrt{\epsilon/24}$, which is independent
of the mass of star. In \citep{safari} this metric is used to
address the Pioneer's anomalous.

%---------------------------------------------------------
\subsection{$f(R)=R\pm\mu^{4}/R$}
 Based on equivalence between $f(R)$
gravity and Brans-Dicke  theory with $\omega=0$, it was argued that
this theory is inconsistent with solar system tests \citep{a5}.
Indeed by this approach the Post-Newtonian parameter is found as
$\gamma_{PPN}=1/2$ while the measurements indicate
$\gamma_{PPN}=1+(2.1\pm 2.3)\times10^{-5}$ \citep{cas}. Also we must
note that using equivalence between $f(R)$ gravity and scalar tensor
gravity one can find models which are consistent with the solar
system tests. This consistency can be made by giving the scalar a
high mass or exploiting the so-called chameleon effect\citep{New1, r9,
r10, r11}. However, when one is using equivalence between $f(R)$
gravity and scalar tensor gravity, the continuity of scalar field or
its equivalent, the Ricci scalar, at the matter boundary is crucial
condition which is not the case in Einstein gravity. But in this
work we don't adopt the continuity of Ricci scalar for solving the
field equations. Instead, we suppose that when $\mu$ tends to zero
we arrive at the Einstein gravity. Thus we find a solution for $1/R$
model which is radically different from other solutions in
\citep{us1, us2}.\par For this model we have
\begin{eqnarray}\label{eq12}
h(R)&=&\pm1/R,
\nonumber
\\
\varphi(R)&=&\mp 1/R^{2},
\end{eqnarray}
where $h(R)$ fulfills the condition $\mathop {\lim }\limits_{R \to
0} \left[ {h(R)/\varphi (R)} \right] = 0$. Solving
Eqs.(\ref{eq7},\ref{eq8}) we obtain
\begin{eqnarray}\label{eq13}
&&R=\mp7\alpha
\mu^{\frac{4}{3}}r^{-\frac{2}{3}},\nonumber \\
&&\frac{\mu^{4}}{R^{2}}=\frac{1}{49\alpha^{2}}\mu^{\frac{4}{3}}r^{\frac{4}{3}},\nonumber\\
&&a=-\frac{2M}{r}\pm\frac{3}{4}\alpha\mu^{\frac{4}{3}}r^{\frac{4}{3}}\label{18},\nonumber \\
&&b=-\frac{2M}{r}\pm\alpha\mu^{\frac{4}{3}}r^{\frac{4}{3}}.
\end{eqnarray}
where $\alpha^{3}=4/147$ and $M$ is a constant. Therefore the metric
for space time is
\begin{eqnarray}\label{eq14}
 ds^2&=&-\left(1-\frac{2M}{r}\pm\frac{3}{4}\alpha\mu^{\frac{4}{3}}r^{\frac{4}{3}}\right)dt^2
 \nonumber
 \\
 &&+\left(1-\frac{2M}{r}\pm\alpha\mu^{\frac{4}{3}}r^{\frac{4}{3}}\right)^{-1}dr^2 + r^2d\Omega^2.
 \end{eqnarray}
 { We can use the isotropic form, by introducing a new radius, $\rho$, which defined as
   \begin{eqnarray}\nonumber
 r=\rho \sqrt{1+\frac{2M}{\rho}\pm\frac{3}{4}\alpha \mu^{\frac{4}{3}} \rho^{\frac{4}{3}}},
 \end{eqnarray}
 and therefore the equivalent metric can be read as
 \begin{eqnarray}\nonumber
 ds^{2}&=&-(1-\frac{2M}{\rho}\pm\frac{3}{4}\alpha \mu^{\frac{4}{3}} \rho^{\frac{4}{3}}) dt^{2}\cr
 &+&
 (1+\frac{2M}{\rho}\pm\frac{3}{4}\alpha \mu^{\frac{4}{3}} \rho^{\frac{4}{3}})(d\rho^{2}+\rho^{2}d\Omega^{2}).
 \end{eqnarray}
 From above metric one can see in the asymptotic behavior,$\mu\rightarrow0$, $\gamma_{PPN}\simeq1$ can be obtained.}\\
From Eq.(\ref{eq13}) it is obvious that in the limit
$\mu\rightarrow0$, $\mu^{4}/R^{2}$ tends to zero, so there is not
singularity in the field equations. Also one can check neglecting
$R$, $G_{\mu\nu}$, and $h(R)/\varphi(R)$ in Eq.(\ref{eq1},
\ref{eq2}) is reasonable.

%---------------------------------------------------------------

\subsection{Interior solution in the $f(R)=R+\mu^{4}/R$ model }
In this section we discuss the interior gravitational field in the
spherically symmetric case of static mass distribution in the
$f(R)=R+\mu^{4}/R$ model where $\mu\rightarrow 0$. So we seek a
spherically symmetric, static solution and we adopt the
metric(\ref{eq5}). In this model we may rewrite field equation
(\ref{eq1}) and trace equation (\ref{eq2}) as
\begin{eqnarray}\label{Mix1}
G_{\mu}^{\nu}&=&\left(\delta_{\mu}^{\nu}R+G_{\mu}^{\nu}+\delta_{\mu}^{\nu}\Box
-\nabla_{\mu}\nabla^{\nu}\right)\frac{\mu^{4}}{R^{2}}+kT_{\mu}^{\nu},\\\nonumber\\
R&=&3\left(R-\Box\right)\frac{\mu^{4}}{R^{2}}-kT.
\end{eqnarray}
From Eq.(\ref{eq2}) it is obvious that as $\mu\rightarrow0$,
$R\rightarrow-kT$, so assuming $\mu^{4}\ll-kT$, in the r.h.s of
Eq.(\ref{Mix1}) we may neglect those terms that contain
$\mu^{4}/R^{2}$. Thus field equations (\ref{Mix1}) reduce to Einstein
equations hence we may write
\begin{eqnarray}\label{Mix2}
G_{\mu}^{\nu}\simeq kT_{\mu}^{\nu}.
\end{eqnarray}
furthermore the conservation equation, ${T^{\mu\nu}}_{;\nu}=0$,
leads to
\begin{eqnarray}\label{Mix3}
p^{\prime}=-\frac{A^{\prime}}{2A}(p+\rho),
\end{eqnarray}
where $p,\rho$ are pressure and density of matter. To obtain metric
components $(A, B)$, we use Eq.(\ref{Mix3})  and $rr$ and $tt$
components of Eq.(\ref{Mix2})
\begin{eqnarray}\label{Mix4}
G_{r}^{r}=\frac{A^{\prime}}{A}\frac{B}{r}+\frac{B-1}{r^{2}}\simeq kp,\\\nonumber\\
G_{r}^{r}=\frac{B^{\prime}}{r}+\frac{B-1}{r^{2}}\simeq -k\rho c^{2}.
\end{eqnarray}
Solving Eq.(\ref{Mix4}) we may write
\begin{eqnarray}\label{Mix5}
B=1-\frac{1}{r}kc^{2}\int^{r}_{0}\rho(x)x^{2}dx+{\cal
O}\left(\frac{\mu^{4}}{k^{2}T^{2}}\right).
\end{eqnarray}

\par From continuity of the metric component
B(r), on the boundary surface $r=r_{0}$ we find
\begin{eqnarray}\label{Mix6}
\frac{kc^{2}}{r_{0}}\int^{r_{0}}_{0}\rho(x)x^{2}dx+\alpha\mu^{\frac{4}{3}}{r_{0}}^{\frac{4}{3}}+{\cal
O}\left(\frac{\mu^{4}}{k^{2}T^{2}}\right)=\frac{2M}{r_{0}},
\end{eqnarray}
where in the above equation we used the empty space solution
Eq.(\ref{eq14}). From the above equation we may determine the
parameter $M$. It is seen that in the $\mu\rightarrow 0$ limit this
constant reduces to the Schwarzschild radius.   Furthermore,
according to cosmological studies $\mu^{2}=10^{-52}m^{-2}$
\citep{c4} so, regarding a typical solar system, in Eq.(\ref{Mix6})
we may neglect terms at order ${\cal
O}\left(\frac{\mu^{4}}{k^{2}T^{2}}\right)$.

From equation (\ref{Mix6}) it is obvious that the physical
interpretation of the parameter $M$ differ from that of general
relativity. Also from the above equation it is clear that in the
$1/R$ gravity the external solution depends on the shape of matter
distribution.

%=======================================================================================================================
\section{Strong Gravity Regime in $R+\mu^4/R$ Model}
 In this section we investigate the existence of strong gravitational field for $f(R)=R+\mu^4/R$ model of $f(R)$ gravity.
We can rewrite the field equation (\ref{eq1}) as
\begin{eqnarray}\label{eq19}
G_\mu^\nu\left(1-\frac{\mu^{4}}{R^{2}}\right)=-\frac{1}{3}\delta_\mu^\nu
R -\nabla_{\mu}\nabla^{\nu} \left(\frac{\mu^{4}}{R^{2}}\right),
\end{eqnarray}
where we have used the trace equation
\begin{eqnarray}\label{eq20}
R=-3[R+\Box]\left(\mu^{4}/R^{2}\right).
\end{eqnarray}
In the above equation we have neglected the energy-momentum tensor
of matter because we investigate the strong gravitational
field around a spherically symmetric distribution of matter.
Adopting the general spherically symmetric metric (\ref{eq5}), we
can rewrite the trace equation (\ref{eq20}) and ($rr$),($tt$)
components of field equation (\ref{eq19}) as
\begin{subequations}
\label{eq21}
\begin{eqnarray}
&-&\left[B\left(\frac{d^2}{dr^2}+\frac{2}{r}\frac{d}{dr}\right)+\frac{1}{2}\left(B^\prime+\frac{BA^\prime}{A}\right)\frac{d}{dr}+R\right]
\nonumber
\\
&\times&\left(\mu^{4}/R^{2}\right)=\frac{R}{3},
\\\nonumber\\
&&\left(\frac{BA^\prime}{rA}+\frac{B-1}{r^2}\right)\left(1-\mu^{4}/R^{2}\right)
\nonumber
\\
&+&\left(B\frac{d^2}{dr^2}+\frac{B^\prime}{2}\frac{d}{dr}\right)\left(\mu^{4}/R^{2}\right)=-\frac{R}{3},
\\\nonumber\\
&&\left(\frac{B^\prime}{r}+\frac{B-1}{r^2}\right)\left(1-\mu^{4}/R^{2}\right)\nonumber\\
&+&\frac{BA^\prime}{2A}\frac{d}{dr}\left(\mu^{4}/R^{2}\right)=-\frac{R}{3},
\end{eqnarray}
\end{subequations}

where($\prime$) denotes derivation with respect to the ($r$). In the
previous section we showed, ($R+\mu^4/R$) model has the week field
solution as
\begin{eqnarray}\label{eq22}
ds^{2} &=&-\left[1-\frac{2M}{r}+\frac{3}{4}\alpha (\mu
r)^{\frac{4}{3}}\right]dt^{2} \nonumber
\\&+&\left[1-\frac{2M}{r}+\alpha (\mu
r)^{\frac{4}{3}}\right]^{-1}dr^{2} +r^{2}d\Omega^{2},
\end{eqnarray}
where $\alpha=(4/147)^{1/3}$ . It is obvious this metric reduces to
Schwarzschild metric in the limit $\mu\rightarrow 0$. Now we seek
the solution of field equation in the limit ($r\rightarrow 2M)$.
Without loss of generality we can assume  $2M=1$. In order to solve
equations (\ref{eq21}) we use some definitions as
\begin{eqnarray}
&&\phi=\gamma/R,\nonumber
\\
&&\gamma=-\mu^{4/3},\nonumber
\\
&& A=1-\frac{1}{r}+\gamma a(r),\nonumber
\\
&&B=1-\frac{1}{r}+\gamma b(r).
\end{eqnarray}
\begin{figure}[t]%============================================================================= Fig 1
\includegraphics[width=0.47\textwidth]{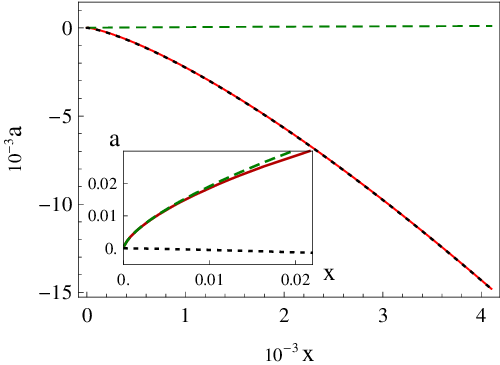}% Here is how to import EPS art
\caption{\label{fig:F1}$a$ against $x$. The red-solid line shows numerical results of Eqs.(32).
The green-dashed line represents approximate solution for $x\ll 1$ (Eq.(34a)) and the black-dotted line is
the approximate solution for $x\gg 1$ (Eq.(36a)).
A close up on the origin of main figure is presented
%to demonstrate the agreement between the approximate solution for $x\ll 1$ and the numerical results
.}
\end{figure}
\begin{figure}[t]%============================================================================= Fig 2
\includegraphics[width=0.47\textwidth]{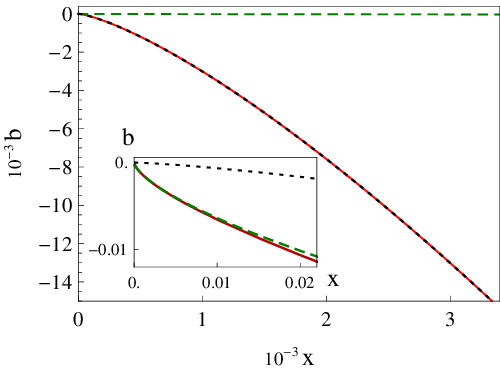}% Here is how to import EPS art
\caption{\label{fig:F2}$b$ against $x$. The red-solid line shows numerical results of Eqs.(32).
The green-dashed line represents approximate solution for $x\ll 1$ (Eq.(34b)) and the black-dotted line is
the approximate solution for $x\gg 1$ (Eq.(36b)).
A close up on the origin of main figure is presented
%to demonstrate the agreement between the approximate solution for $x\ll 1$ and the numerical results
.}
\end{figure}
\begin{figure}[t]%============================================================================= Fig 3
\includegraphics[width=0.47\textwidth]{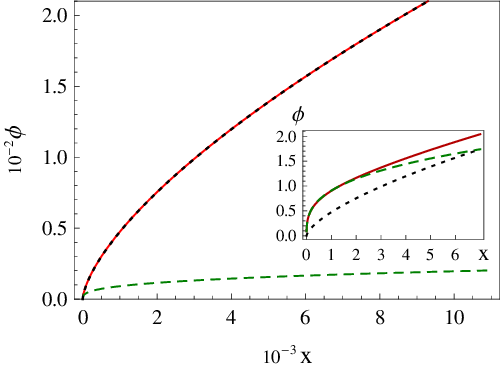}% Here is how to import EPS art
\caption{\label{fig:F3}$\varphi$ versus $x$. The red-solid line shows numerical results of Eq.(32a).
The green-dashed line represents approximate solution for $x\ll 1$ (Eq.(34c)) and the black-dotted line is
the approximate solution for $x\gg 1$ (Eq.(36c)).
A close up on the origin of main figure is presented
%to demonstrate the agreement between the approximate solution for $x\ll 1$ and the numerical results
.}
\end{figure}
Because we seek the solution in the limit $r\rightarrow 1 $, we may
define a new variable as $x=r-1$. Using these definitions we can
rewrite Eq.(\ref{eq21}) as
\begin{subequations}
\label{eq23}
\begin{eqnarray}
&&\gamma\left(b\frac{d}{dx}+\frac{2b}{x+1}+\frac{b^\prime+a^\prime}{2}+\frac{(x+1)(b-a)}{2(x+\gamma
a(x+1))}  \right.\nonumber\\&&\left.
\left(\frac{1}{(x+1)^{2}}+\gamma
a^\prime\right)\right)\frac{d\phi^2}{dx}=\left(\frac{1}{3}-\gamma\phi^{2}\right)\frac{1}{\phi}\nonumber\\&&-\left(\frac{x}{x+1}\frac{d^2}{dx^2}+\frac{2x+1}{(x+1)^2}\frac{d}{dx}\right)\phi^{2}
\end{eqnarray}
\begin{eqnarray}
&&\left(\frac{x}{x+1}\frac{d^2}{dx^2}+\frac{1}{2(x+1)^2}\frac{d}{dx}+
\gamma\left(b\frac{d^2}{dx^2}+\frac{b^\prime}{2}\frac{d}{dx}\right)\right)\phi^{2}\nonumber\\
&&=\frac{1}{3\phi}+\left(\frac{b}{(x+1)^2}+\frac{a^\prime}{x+1}+\frac{b-a}{x+\gamma
a(x+1)}\right.
\nonumber\\
&&\times \left.\left(\frac{1}{(x+1)^{2}}+\gamma a^\prime\right)\right)(1+\gamma\phi^2)
\end{eqnarray}
\begin{eqnarray}
&&\frac{1}{2}\left(1+\gamma \frac{(x+1)(b-a)}{x+\gamma
a(x+1)}\right)\left(\frac{1}{(x+1)^{2}}+\gamma
a^\prime\right)\frac{d\phi^2}{dx}\nonumber\\&&=\frac{1}{3\phi}+\left(\frac{b}{(x+1)^2}+\frac{b^{\prime}}{x+1}\right)\left(1+\gamma\phi^{2}\right),
\end{eqnarray}
\end{subequations}
where ($\prime$) denotes derivation with respect to the ($x$). For
the limit $\mu\rightarrow 0$, in the above equations we suppose that we
can neglect terms containing $\gamma$ . After solving equations we
check this assumption. By neglecting these terms, equations \ref{eq23}
can be rewritten as
\begin{subequations}
\label{eq24}
\begin{eqnarray}
&&\frac{1}{3\phi}=\left(\frac{x}{x+1}\frac{d^2}{dx^2}+\frac{2x+1}{(x+1)^2}\frac{d}{dx}\right)\phi^{2}
\end{eqnarray}
\begin{eqnarray}
&&\frac{b}{(x+1)^2}+\frac{a^\prime}{x+1}+\frac{b-a}{x(x+1)^{2}}=-\frac{1}{3\phi}
\nonumber\\
&&+\left(\frac{x}{x+1}\frac{d^2}{dx^2}+\frac{1}{2(x+1)^2}\frac{d}{dx}\right)\phi^{2}
\end{eqnarray}
\begin{eqnarray}
&&\frac{1}{2}\frac{1}{(x+1)^{2}}\frac{d\phi^2}{dx}=\frac{1}{3\phi}+\frac{b}{(x+1)^2}+\frac{b^{\prime}}{x+1}.
\end{eqnarray}
\end{subequations}
In the limit $x\ll1$, solutions of Eq. (\ref{eq24}) are
\begin{subequations}
\label{eq25}
\begin{eqnarray}
&&a_{0}=\frac{3}{8}\left(\frac{4}{3}\right)^{1/3}x^{2/3},\\
&&b_{0}=-\frac{1}{8}\left(\frac{4}{3}\right)^{1/3}x^{2/3},\\
&&\phi_{0}=\left(\frac{3}{4}\right)^{1/3}x^{1/3}.
\end{eqnarray}
\end{subequations}
Thus we
obtain the metric for $x\ll1$ as
\begin{eqnarray}\label{eq27}
ds^{2}
&=&-\left(1-\frac{1}{r}-\frac{3}{8}\left(\frac{4}{3}\right)^{1/3}\mu^{4/3}(r-1)^{2/3}\right)dt^{2}
\nonumber\\
&+&\left(1-\frac{1}{r}+\frac{1}{8}\left(\frac{4}{3}\right)^{1/3}\mu^{4/3}(r-1)^{2/3}\right)dr^{2}
\nonumber\\&+&r^{2}d\Omega^{2}.
\end{eqnarray}
Furthermore, for $x\gg1$, we can obtain the solutions of equations
(\ref{eq24}) as
\begin{subequations}
\label{eq26}
\begin{eqnarray}
a_{\infty}&=&-\frac{3}{4}\alpha x^{4/3},
\\
b_{\infty}&=&-\alpha
x^{4/3},
\\
\phi_{\infty}&=&\frac{1}{7\alpha}x^{2/3},
\end{eqnarray}
\end{subequations}
which are in agreement with week field limit (\ref{eq22}).
Now we can check the validity of our assumption. Considering the
solutions (\ref{eq26}), shows that neglecting terms containing
$\gamma$ in Eqs. (\ref{eq23}) is valid only for
$x\gg\mid\gamma^{3}\mid$ or $x\gg\mu^4$. Hence the metric
(\ref{eq25}) is solution of field equations in the range of
$\mu^{4}\ll x\ll 1$. By performing a conformal transformation and
changing coordinate we can see the strong field solution (\ref{eq27}) is
\begin{eqnarray}\label{eq28}
&&ds^{2}=\nonumber\\ &&-\left(1-\frac{2M}{r}-\frac{3}{8}\left(\frac{4}{3}\right)^{1/3}(2M\mu)^{4/3}(\frac{r}{2M}-1)^{2/3}\right)dt^{2}
\nonumber\\
&&+\left(1-\frac{2M}{r}+\frac{1}{8}\left(\frac{4}{3}\right)^{1/3}(2M\mu)^{4/3}(\frac{r}{2M}-1)^{2/3}\right)dr^{2}
\nonumber\\
&&+r^{2}d\Omega^{2}\nonumber,
\end{eqnarray}
which is valid in the range of $(2M\mu)^{4} \ll r/2M-1 \ll 1$  and
farther where $r\gg 2M$, the metric of space time can be
approximated by the metric (\ref{eq22}).  Furthermore, we have
solved field equations (\ref{eq23})numerically and presented the results in figures \ref{fig:F1}, \ref{fig:F2}, and \ref{fig:F3}.
The plots show that the numerical results are in agreement with the analytical solutions (\ref{eq25},\ref{eq26}) in their region of validity.

\section{Discussion}
We studied spherically symmetric solution of $f(R)$ gravity. At first a new approach for investigating
the weak field limit of vacuum field equations in $f(R)$ gravity was introduced. Our results for the weak field limit of some studied f(R) models are in
agreement with their known solutions. We solved the field equations for $f(R)=R+\mu ^4/R$ gravity at weak field limit and obtained a solution which differs
slightly from the schwarzschild metric. { Our results are against the arguments that $f(R)$ models are ill defined because of the equivalence of $f(R)$
gravity and Brans-Dicke gravity with $\omega_{BD}=0$ which leads to $\gamma_{PPN}=1/2$. In fact our results are in agreement with the recent article of
Capozziello et al. \citep{rr14}, in which they have studied Newtonian limit of the $f(R)$ gravity by considering that fourth order gravity models are dynamically
equivalent to the O'Hanlon lagrangian and they have shown fourth order gravity models can not be ruled out only on the base of analogy with Brans-Dicke gravity
with $\omega_{BD}=0$.}
 Moreover, regarding the results for the weak field limit, we investigated the strong field regime for this model and  showed that if
 $(r-2M)/(2M)^{5}\gg\mu^{4}$, where $r$ and $2M$ are
radius and Schwarzschild radius in the Schwarzschild coordinate
respectively, the gravitational field is a perturbed Schwarzschild
metric even in strong gravity regime.
finally we solved the master equations numerically by setting the initial value conditions using the analytical answers of the strong gravity region. In figures
(\ref{fig:F1}) and (\ref{fig:F2}) we plotted the analytical and numerical solutions of the components of the metric, $a$ and $b$, versus radius in two weak and
strong gravity region. It is seen that in the strong region (the close up part) the relevant analytical answer and the numerical solution are agree together
while the analytical weak field approximation solution deviates from the numerical solution. The close up part of figures show that with increasing the radius
and going to the weak filed region, the  analytical solutions of strong filed approximation and numerical answers get separated from each other, and at last in
the weak field region, i.e. $x \gg 1$, the analytical weak field answers coincide with the numerical solution, while the answers for the strong gravity region
has a grate deviation from the numerical results in this region.

\end{document}